\documentstyle[epsfig]{aipproc}
\begin{document}
\input psfig.tex
\title{Optical Observations of Type II Supernovae}

\author{Alexei V. Filippenko}
\address{Department of Astronomy, University of California,
Berkeley, CA 94720-3411}
\maketitle

\begin{abstract} 
I present an overview of optical observations (mostly spectra)
of Type II supernovae. SNe~II are defined by the presence of hydrogen, and
exhibit a very wide variety of properties. SNe~II-L tend to show evidence of
late-time interaction with circumstellar material.  SNe~IIn are distinguished
by relatively narrow emission lines with little or no P-Cygni absorption
component and (quite often) slowly declining light curves; they probably have
unusually dense circumstellar gas with which the ejecta interact. Some SNe~IIn,
however, might not be genuine SNe, but rather are super-outbursts of luminous
blue variables.  The progenitors of SNe~IIb contain only a low-mass skin of
hydrogen; their spectra gradually evolve to resemble those of SNe~Ib. Limited
spectropolarimetry thus far indicates large asymmetries in the ejecta of
SNe~IIn, but much smaller ones in SNe~II-P. There is intriguing, but still
inconclusive, evidence that some peculiar SNe~IIn might be associated with
gamma-ray bursts. SNe~II-P are useful for cosmological distance determinations
with the Expanding Photosphere Method, which is independent of the 
Cepheid distance scale.
\end{abstract}

\section*{ INTRODUCTION}

   Supernovae (SNe) occur in several spectroscopically distinct varieties; see
reference \cite{avf97}, for example.  Type I SNe are defined by the absence of
obvious hydrogen in their optical spectra, except for possible contamination
from superposed H~II regions.  SNe~II all prominently exhibit hydrogen in their
spectra, yet the strength and profile of the H$\alpha$ line vary widely among
these objects.

   The early-time ($t \approx 1$ week past maximum brightness) spectra of SNe
are illustrated in Figure 1. [Unless otherwise noted, the optical spectra
illustrated here were obtained by my group, primarily with the 3-m Shane
reflector at Lick Observatory. When referring to phase of evolution, the
variables $t$ and $\tau$ denote time since {\it maximum brightness} (usually in
the $B$ passband) and time since {\it explosion}, respectively.]  The lines are
broad due to the high velocities of the ejecta, and most of them have P-Cygni
profiles formed by resonant scattering above the photosphere. SNe~Ia are
characterized by a deep absorption trough around 6150~\AA\ produced by
blueshifted Si~II $\lambda$6355. Members of the Ib and Ic subclasses do not
show this line. The presence of moderately strong optical He~I lines,
especially He~I $\lambda$5876, distinguishes SNe~Ib from SNe~Ic.

\bigskip

\hbox{
\hskip +.7truein
\psfig{figure=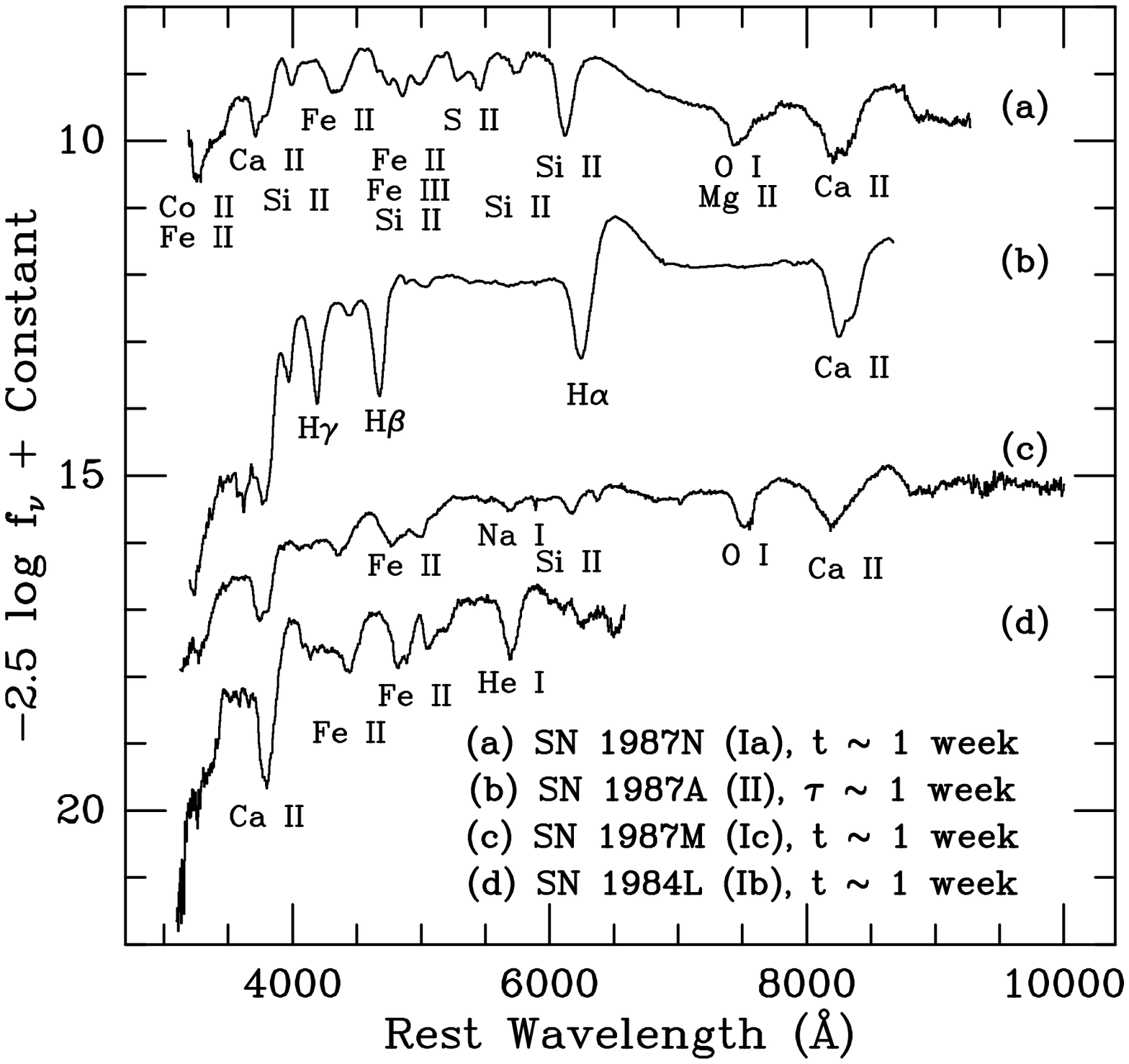,height=4.5truein,width=4.5truein,angle=0}
}
\noindent
{\it Figure 1:} Early-time spectra of SNe, showing the main subtypes.\\
\medskip

   The late-time ($t \gtrsim 4$ months) optical spectra of SNe provide
additional constraints on the classification scheme (Figure 2). SNe~Ia show
blends of dozens of Fe emission lines, mixed with some Co lines.  SNe~Ib and
Ic, on the other hand, have relatively unblended emission lines of
intermediate-mass elements such as O and Ca. At this phase, SNe~II are
dominated by the strong H$\alpha$ emission line; in other respects, most of
them spectroscopically resemble SNe~Ib and Ic, but with narrower emission
lines. The late-time spectra of SNe~II show substantial heterogeneity, as do
the early-time spectra.

  To a first approximation, the light curves of SNe~I are all broadly similar
\cite{lei91a}, while those of SNe~II exhibit much dispersion \cite{pat93}.  It
is useful to subdivide the majority of early-time light curves of SNe~II into
two relatively distinct subclasses \cite{bar79,dog85}.  The light curves of
SNe~II-L (``linear'') generally resemble those of SNe~I, with a steep decline
after maximum brightness followed by a slower exponential tail. In contrast,
SNe~II-P (``plateau") remain within $\sim 1$ mag of maximum brightness for an
extended period. The peak absolute magnitudes of SNe~II-P show a very wide
dispersion \cite{you89}, almost certainly due to differences in the radii of
the progenitor stars. The light curve of SN 1987A, albeit unusual, was
generically related to those of SNe~II-P; the initial peak was very low because
the progenitor was a blue supergiant, much smaller than a red supergiant
\cite{arn89}. The remainder of this review concentrates on SNe~II.  

\bigskip

\hbox{
\hskip +.7truein
\psfig{figure=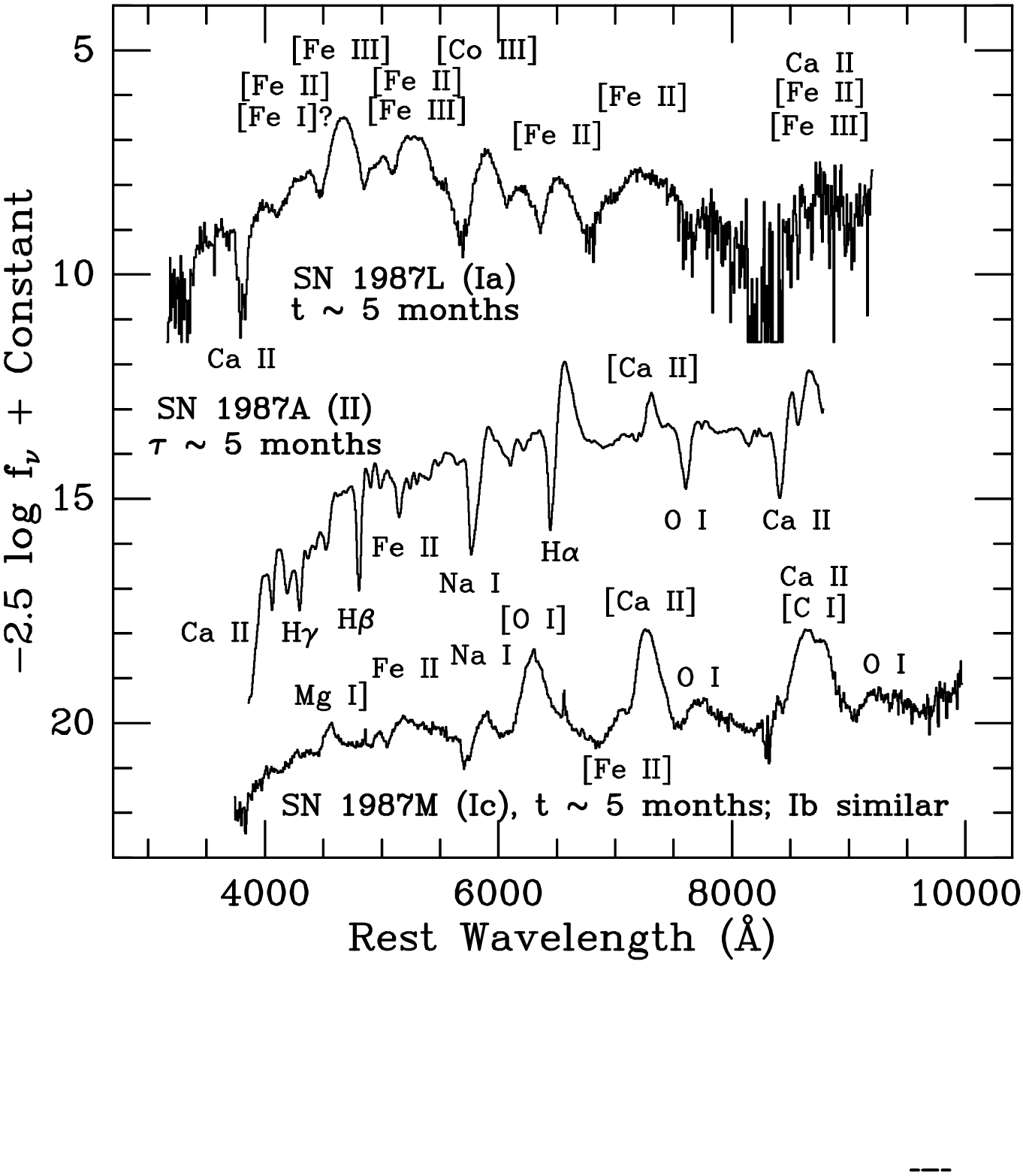,height=4.5truein,width=4.5truein,angle=0}
}

\noindent
{\it Figure 2:} Late-time spectra of SNe.  At even later phases, SN
1987A was dominated by strong emission lines of H$\alpha$, [O~I], [Ca~II], and
the Ca~II near-infrared triplet.
\medskip

\section*{ SUBCLASSES OF TYPE II SUPERNOVAE}

  Most SNe~II-P seem to have a relatively well-defined spectral development, as
shown in Figure 3 for SN 1992H (see also reference \cite{clo96}). At early
times the spectrum is nearly featureless and very blue, indicating a high color
temperature ($\gtrsim$ 10,000~K). He~I $\lambda$5876 with a P-Cygni profile
is sometimes visible. The temperature rapidly decreases with time, reaching
$\sim 5000$~K after a few weeks, as expected from the adiabatic expansion and
associated cooling of the ejecta. It remains roughly constant at this value
during the plateau (the photospheric phase), while the hydrogen recombination
wave moves through the massive ($\sim 10~M_\odot$) hydrogen ejecta and releases
the energy deposited by the shock. At this stage strong Balmer lines and Ca~II
H\&K with well-developed P-Cygni profiles appear, as do weaker lines of Fe~II,
Sc~II, and other iron-group elements. The spectrum gradually takes on a nebular
appearance as the light curve drops to the late-time tail; the continuum fades,
but H$\alpha$ becomes very strong, and prominent emission lines of [O~I],
[Ca~II], and Ca~II also appear.

\bigskip

\hbox{
\hskip +1truein
\psfig{figure=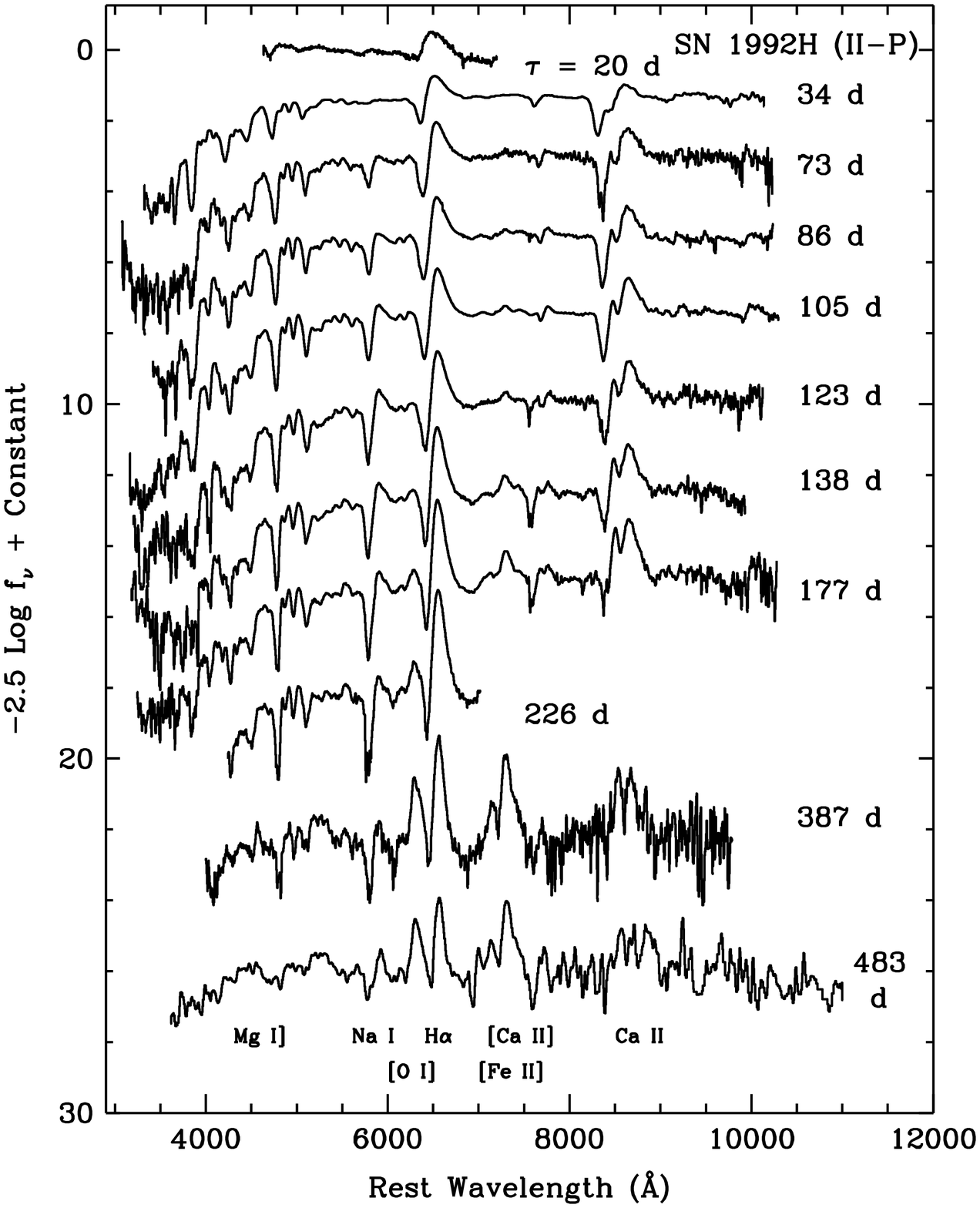,height=5truein,width=4truein,angle=0}
}

\noindent
{\it Figure 3:} Montage of spectra of SN 1992H in NGC 5377. Epochs
(days) are given relative to the estimated time of explosion,
February 8, 1992.
\medskip

   Few SNe~II-L have been observed in as much detail as SNe~II-P.  Figure 4
shows the spectral development of SN 1979C \cite{bra81}, an unusually luminous
member of this subclass. Near maximum brightness the spectrum is very blue and
almost featureless, with a slight hint of H$\alpha$ emission. A week later,
H$\alpha$ emission is more easily discernible, and low-contrast P-Cygni
profiles of Na~I, H$\beta$, and Fe~II have appeared. By $t \approx 1$ month,
the H$\alpha$ emission line is very strong but still devoid of an absorption
component, while the other features clearly have P-Cygni profiles. Strong,
broad H$\alpha$ emission dominates the spectrum at $t \approx 7$ months, and
[O~I] $\lambda\lambda$6300, 6364 emission is also present.  Several authors
\cite{whe90,avf91a,sch96} have speculated that the absence of H$\alpha$
absorption spectroscopically differentiates SNe~II-L from SNe~II-P, but the
small size of the sample of well-observed objects precluded definitive
conclusions.

\bigskip

\hbox{
\hskip +1truein
\psfig{figure=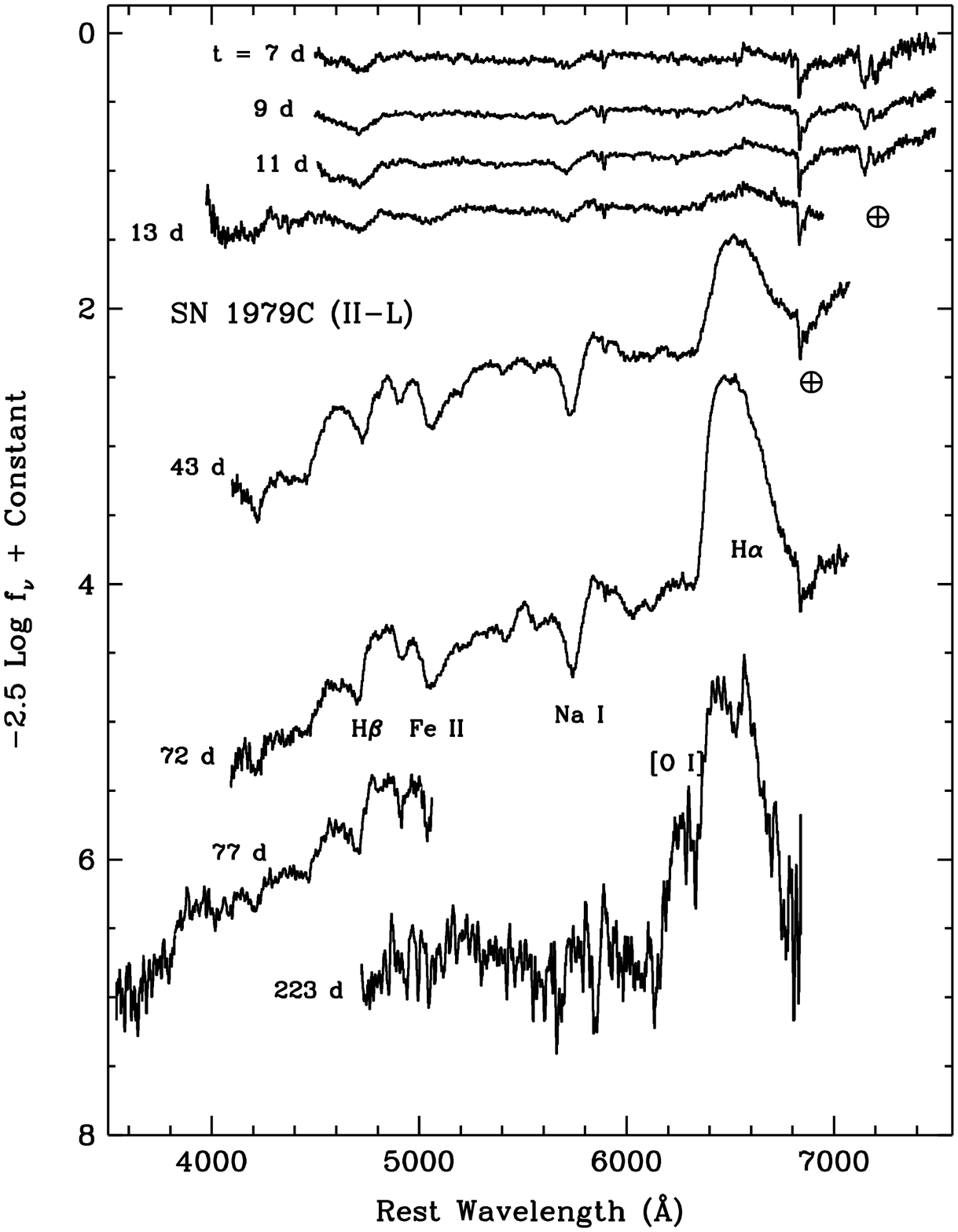,height=5truein,width=4truein,angle=0}
}

\noindent
{\it Figure 4:} Montage of spectra of SN 1979C in NGC 4321, from
reference \cite{bra81}; reproduced with permission. Epochs
(days) are given relative to the date of maximum brightness,
April 15, 1979.
\medskip

   The progenitors of SNe~II-L are generally believed to have relatively
low-mass hydrogen envelopes (a few $M_\odot$); otherwise, they would exhibit
distinct plateaus, as do SNe~II-P. On the other hand, they may have more
circumstellar gas than do SNe~II-P, and this could give rise to the
emission-line dominated spectra. They are often radio sources \cite{sra90};
moreover, the ultraviolet excess (at $\lambda \lesssim 1600$~\AA) seen in SNe
1979C and 1980K may be produced by inverse Compton scattering of photospheric
radiation by high-speed electrons in shock-heated ($T \approx 10^9$~K)
circumstellar material \cite{fra82,fra84}. Finally, the light curves of some
SNe~II-L reveal an extra source of energy: after declining exponentially for
several years, the H$\alpha$ flux of SN 1980K reached a steady level, showing
little if any decline thereafter \cite{uom86,lei91b}.  The excess almost
certainly comes from the kinetic energy of the ejecta being thermalized and
radiated due to an interaction with circumstellar matter \cite{che90,lei94}.

   The very late-time optical recovery of SNe 1979C and 1980K
\cite{lei91b,fes95,fes99} and other SNe~II-L supports the idea of ejecta
interacting with circumstellar material.  The spectra consist of a few strong,
broad emission lines such as H$\alpha$, [O~I] $\lambda\lambda$6300, 6364, and
[O~III] $\lambda\lambda$4959, 5007.  A {\it Hubble Space Telescope (HST)}
ultraviolet spectrum of SN 1979C reveals some prominent, double-peaked emission
lines with the blue peak substantially stronger than the red, suggesting dust
extinction within the expanding ejecta \cite{fes99}. The data show general
agreement with the emission lines expected from circumstellar interaction
\cite{che94}, but the specific models that are available show several
differences with the observations. For example, we find higher electron
densities ($10^5$ to $10^7$ cm$^{-3}$), resulting in stronger collisional
de-excitation than assumed in the models. These differences can be used to
further constrain the nature of the progenitor star. Note that based on
photometry of the stellar populations in the environment of SN 1979C (from {\it
HST} images), the progenitor of the SN was at most 10 million years years old,
so its initial mass was probably 17--18~$M_\odot$ \cite{van99a}.

   During the past decade, there has been the gradual emergence of a new,
distinct subclass of SNe~II \cite{avf91a,avf91b,sch90,lei94} whose ejecta are
believed to be {\it strongly} interacting with dense circumstellar gas, even at
early times (unlike SNe~II-L). The derived mass-loss rates for the progenitors
can exceed $10^{-4} M_\odot$ yr$^{-1}$ \cite{chu94}.  In these objects, the
broad absorption components of all lines are weak or absent throughout their
evolution.  Instead, their spectra are dominated by strong emission lines, most
notably H$\alpha$, having a complex but relatively narrow profile. Although the
details differ among objects, H$\alpha$ typically exhibits a very narrow
component (FWHM $\lesssim 200$ km s$^{-1}$) superposed on a base of
intermediate width (FWHM $\approx$ 1000--2000 km s$^{-1}$; sometimes a very
broad component (FWHM $\approx$ 5000--10,000 km s$^{-1}$) is also present. This
subclass was christened ``Type IIn" \cite{sch90}, the ``n" denoting ``narrow"
to emphasize the presence of the intermediate-width or very narrow emission
components. Representative spectra of five SNe~IIn are shown in Figure 5, with
two epochs for SN 1994Y.

   The early-time continua of SNe~IIn tend to be bluer than normal.
Occasionally He~I emission lines are present in the first few spectra (e.g., SN
1994Y in Figure 5). Very narrow Balmer absorption lines are visible in the
early-time spectra of some of these objects, often with corresponding Fe~II,
Ca~II, O~I, or Na~I absorption as well (e.g., SNe 1994W and 1994ak in Figure
5). Some of them are unusually luminous at maximum brightness, and they
generally fade quite slowly, at least at early times. The equivalent width of
the intermediate H$\alpha$ component can grow to astoundingly high values at
late times. The great diversity in the observed characteristics of SNe~IIn
provides clues to the various degrees and forms of mass loss late in the lives
of massive stars.

\bigskip

\hbox{
\hskip +0.6truein
\psfig{figure=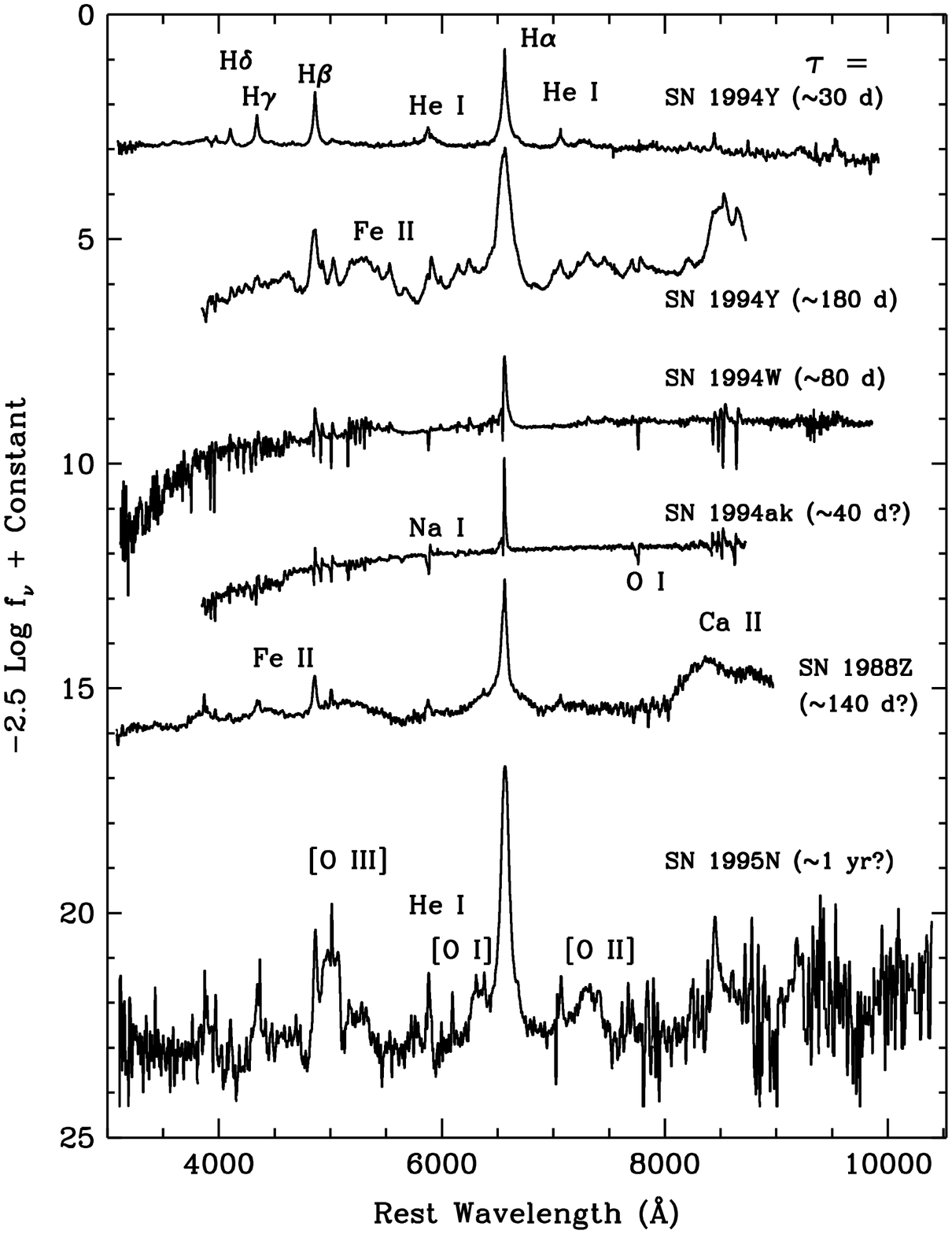,height=5.5truein,width=4.5truein,angle=0}
}
\bigskip
\noindent
{\it Figure 5:} Montage of spectra of SNe~IIn. Epochs are given relative 
to the estimated dates of explosion.\\

\bigskip

\section*{ TYPE II SUPERNOVA IMPOSTORS?}

  The peculiar SN~IIn 1961V (``Type V" according to Zwicky \cite{zwi65}) had
probably the most bizarre light curve ever recorded. (SN 1954J, also known as
``Variable 12" in NGC 2403, was similar \cite{hum94}.)  Its progenitor was a
very luminous star, visible in many photographs of the host galaxy (NGC 1058)
prior to the explosion. Perhaps SN 1961V was not a genuine supernova (defined
to be the violent destruction of a star at the end of its life), but rather the
super-outburst of a luminous blue variable such as $\eta$ Carinae
\cite{goo89,avf95}.

   A related object may have been SN~IIn 1997bs, the first SN discovered in the
Lick Observatory Supernova Search (LOSS) that we are conducting with the 0.75-m
Katzman Automatic Imaging Telescope (KAIT) at Lick Observatory \cite{wli00}.
Its spectrum was peculiar (Figure 6), consisting of narrow Balmer and Fe~II
emission lines superposed on a featureless continuum. Its progenitor was
discovered in an {\it HST} archival image of the host galaxy \cite{van99b}. It
is a very luminous star ($M_V \approx -7.4$ mag), and it didn't brighten as
much as expected for a SN explosion ($M_V \approx -13$ at maximum). These data
suggest that SN 1997bs may have been like SN 1961V --- that is, a supernova
impostor. The real test will be whether the star is still visible in future
{\it HST} images obtained years after the outburst.

\bigskip

\hbox{
\hskip +0.3truein
\psfig{figure=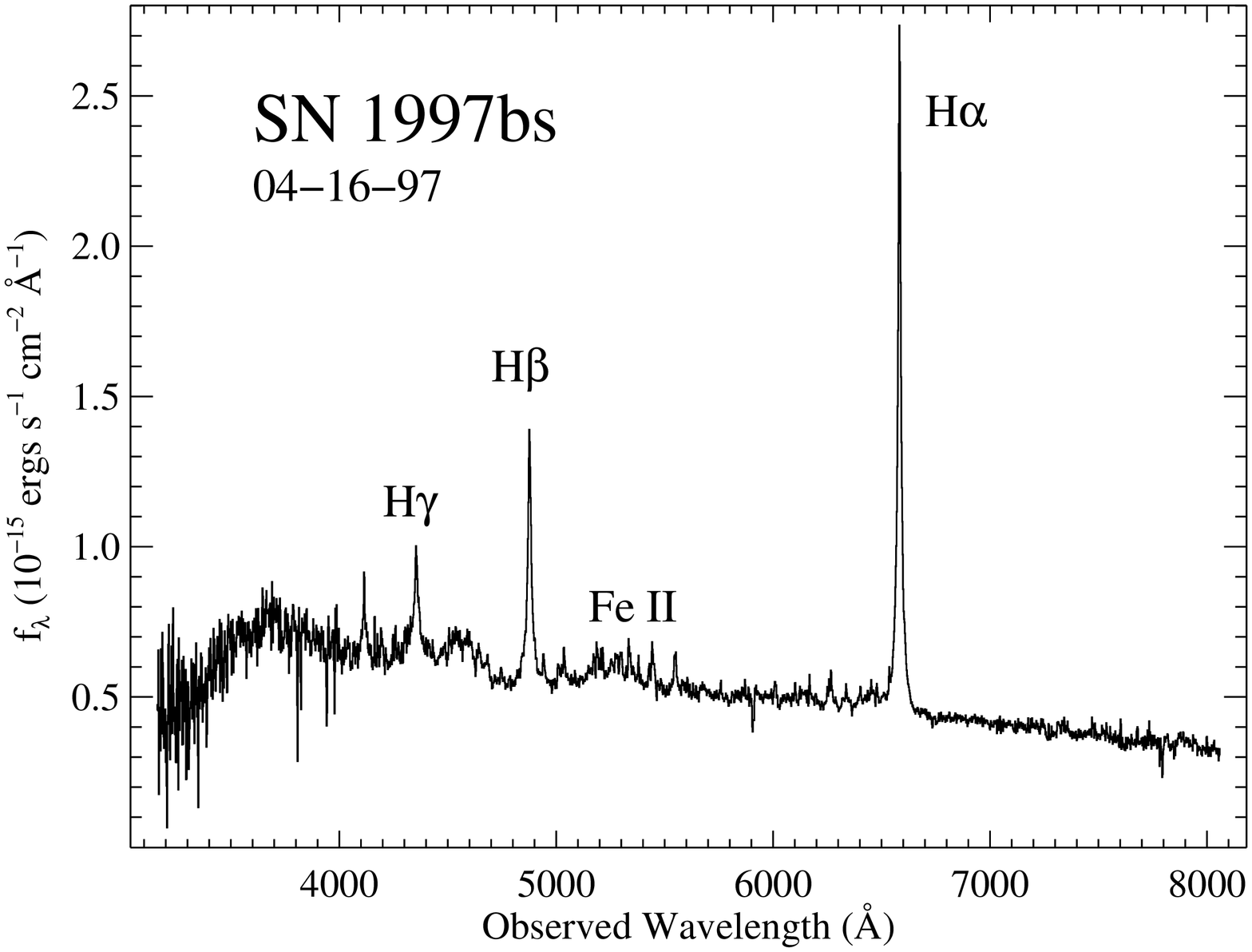,height=5truein,width=3truein,angle=90}
}
\bigskip
\noindent
{\it Figure 6:} Spectrum of SN 1997bs, obtained on April 16, 1997 UT.\\

\bigskip

\section*{ LINKS BETWEEN TYPE II AND TYPE Ib/Ic SUPERNOVAE}

   Filippenko \cite{avf88} discussed the case of SN 1987K, which appeared to be
a link between SNe~II and SNe~Ib. Near maximum brightness, it was undoubtedly a
SN~II, but with rather weak photospheric Balmer and Ca~II lines. Many months
after maximum brightness, its spectrum was essentially that of a SN~Ib. The
simplest interpretation is that SN 1987K had a meager hydrogen atmosphere at
the time it exploded; it would naturally masquerade as a SN~II for a while, and
as the expanding ejecta thinned out the spectrum would become dominated by
emission from deeper and denser layers. The progenitor was probably a star
that, prior to exploding via iron core collapse, lost almost all of its
hydrogen envelope either through mass transfer onto a companion or as a result
of stellar winds.  Such SNe were dubbed ``SNe~IIb" by Woosley et
al. \cite{woo87}, who had proposed a similar preliminary model for SN 1987A
before it was known to have a massive hydrogen envelope.

\bigskip

\hbox{
\hskip +0.3truein
\psfig{figure=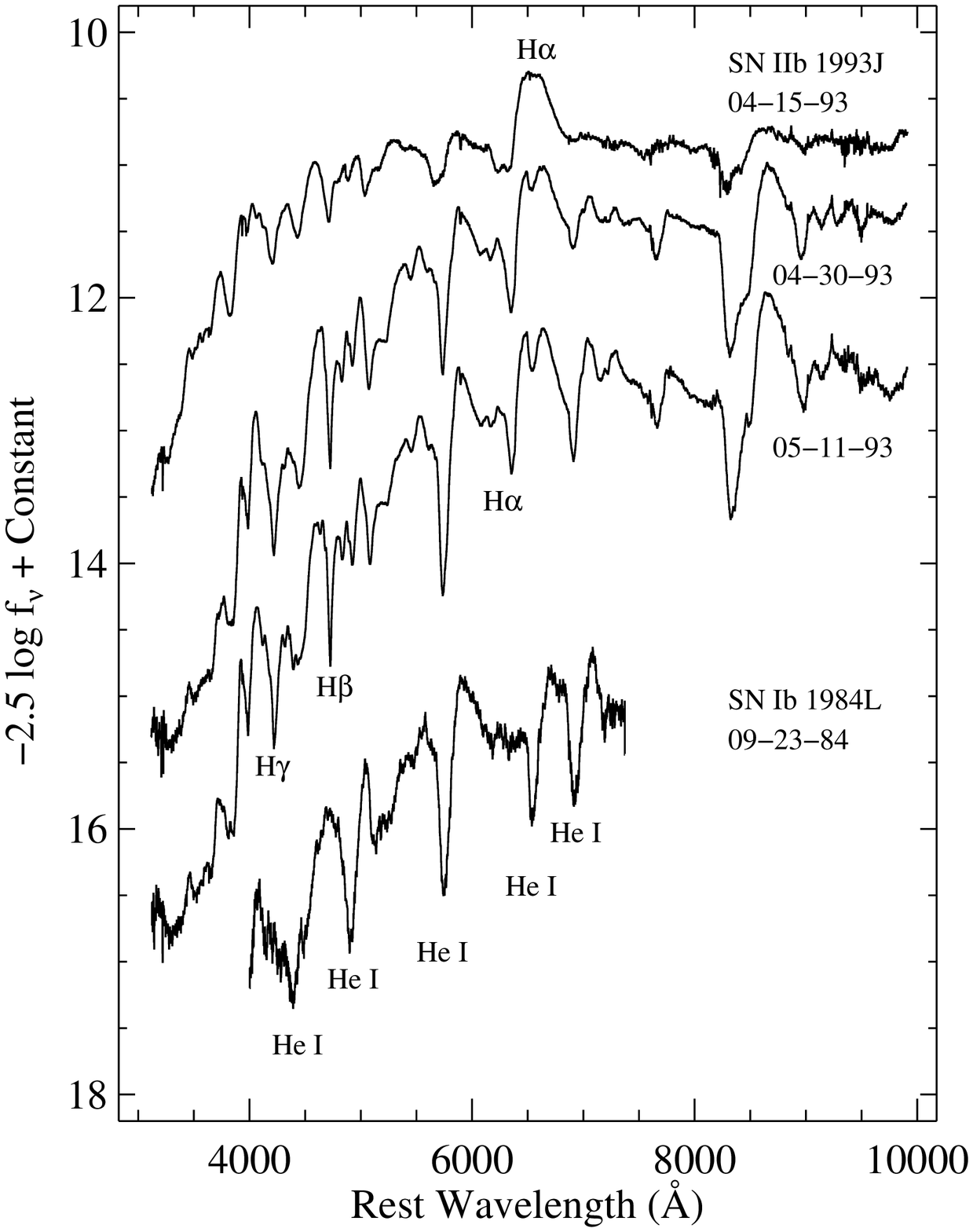,height=5truein,width=5truein,angle=0}
}
\bigskip
\noindent
{\it Figure 7:} Early-time spectral evolution of SN 1993J. A comparison
with the Type Ib SN 1984L is shown at bottom, demonstrating the
presence of He~I lines in SN 1993J. The explosion date was
March 27.5, 1993.\\

\bigskip

   The data for SN 1987K (especially its light curve) were rather sparse,
making it difficult to model in detail.  Fortunately, the Type II SN 1993J in
NGC 3031 (M81) came to the rescue, and was studied in greater detail than any
supernova since SN 1987A \cite{whe96}.  Its light curves \cite{ric96} and
spectra \cite{avf93,avf94,mat00} amply supported the hypothesis that the
progenitor of SN 1993J probably had a low-mass (0.1--0.6~$M_\odot$) hydrogen
envelope above a $\sim 4~M_\odot$ He core \cite{nom93,pod93,woo94}. Figure 7
shows several early-time spectra of SN 1993J, showing the emergence of He~I
features typical of SNe~Ib. Considerably later (Figure 8), the H$\alpha$
emission nearly disappeared, and the spectral resemblance to SNe~Ib was
strong.  The general consensus is that its initial mass was $\sim
15~M_\odot$. A star of such low mass cannot shed nearly its entire hydrogen
envelope without the assistance of a companion star. Thus, the progenitor of SN
1993J probably lost most of its hydrogen through mass transfer to a bound
companion 3--20~AU away. In addition, part of the gas may have been lost from
the system.  Had the progenitor lost essentially {\it all} of its hydrogen
prior to exploding, it would have had the optical characteristics of
SNe~Ib. There is now little doubt that most SNe~Ib, and probably
SNe~Ic as well, result from core collapse in stripped, massive stars, rather
than from the thermonuclear runaway of white dwarfs.

   SN 1993J held several more surprises. Observations at radio \cite{van94} and
X-ray \cite{suz95} wavelengths revealed that the ejecta are interacting with
relatively dense circumstellar material \cite{fra96}, probably ejected from the
system during the course of its pre-SN evolution. Optical evidence for this
interaction also began emerging at $\tau \gtrsim 10$ months: the H$\alpha$
emission line grew in relative prominence, and by $\tau \approx 14$ months it
had become the dominant line in the spectrum \cite{avf94,pat95,fin95},
consistent with models \cite{che94}.  Its profile was very broad (FWHM
$\approx$ 17,000 km s$^{-1}$; Figure 8) and had a relatively flat top, but with
prominent peaks and valleys whose likely origin is Rayleigh-Taylor
instabilities in the cool, dense shell of gas behind the reverse shock
\cite{che92}. Radio VLBI measurements show that the ejecta are circularly
symmetric, but with significant emission asymmetries \cite{mar95}, possibly
consistent with the asymmetric H$\alpha$ profile seen in some of the spectra
\cite{avf94}.

\bigskip

\hbox{
\hskip +0.3truein
\psfig{figure=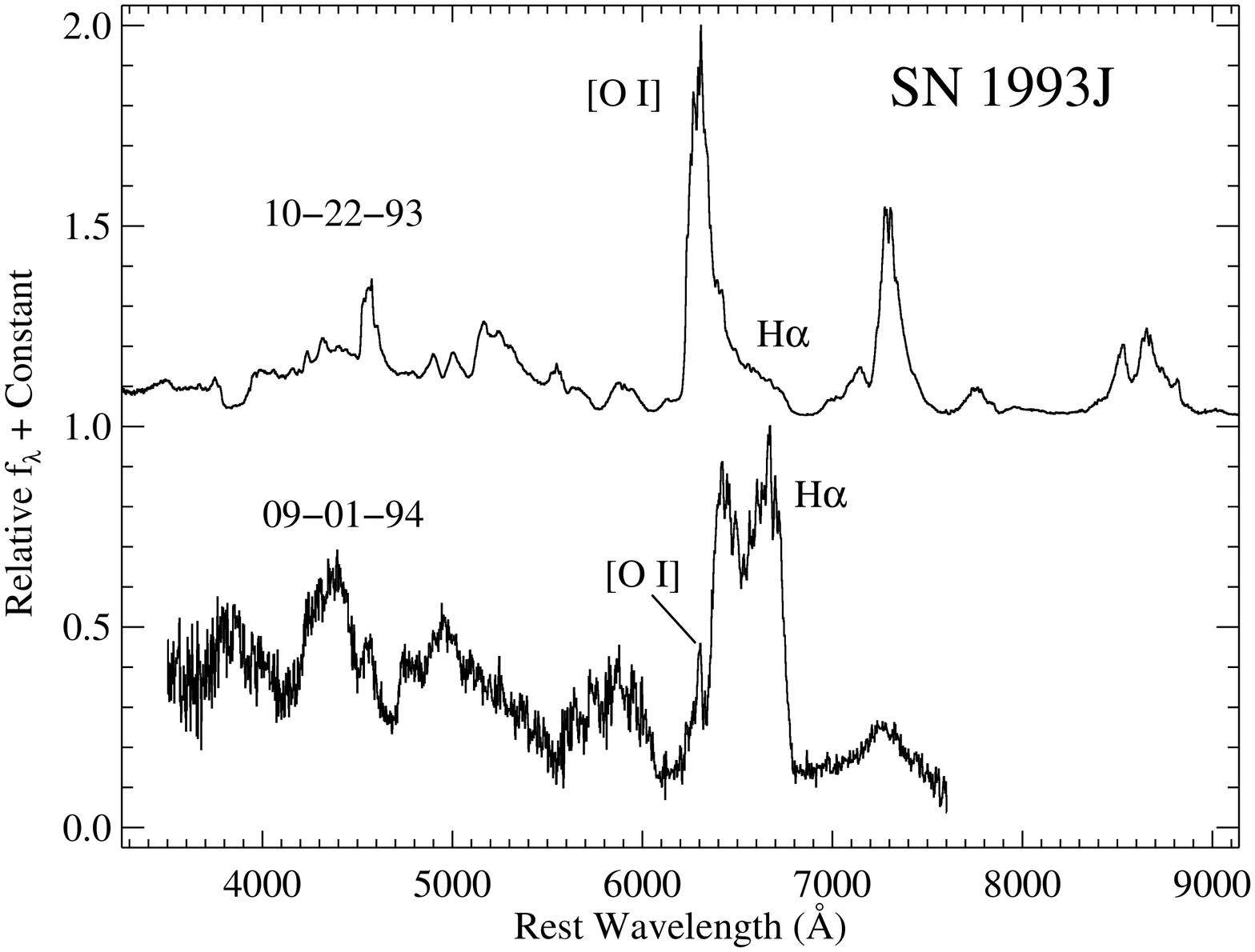,height=5truein,width=3truein,angle=90}
}
\bigskip
\noindent
{\it Figure 8:} In the {\it top} spectrum, which shows SN 1993J about 7
months after the explosion, H$\alpha$ emission is very weak; the
resemblance to spectra of SNe~Ib is striking. A year later ({\it bottom}), 
however, H$\alpha$ was once again the dominant feature
in the spectrum (which was scaled for display purposes).\\

\bigskip

\section*{ SPECTROPOLARIMETRY OF TYPE II SUPERNOVAE}

   Spectropolarimetry of SNe can be used to probe their geometry
\cite{leo00a}. The basic question is whether SNe are round. Such work is
important for a full understanding of the physics of SN explosions and can
provide information on the circumstellar environment of SNe. We have obtained
spectropolarimetry of one object from each of the major SN types and subtypes,
generally with the Keck-II 10-m telescope.

\bigskip

\hbox{
\hskip +1truein
\psfig{figure=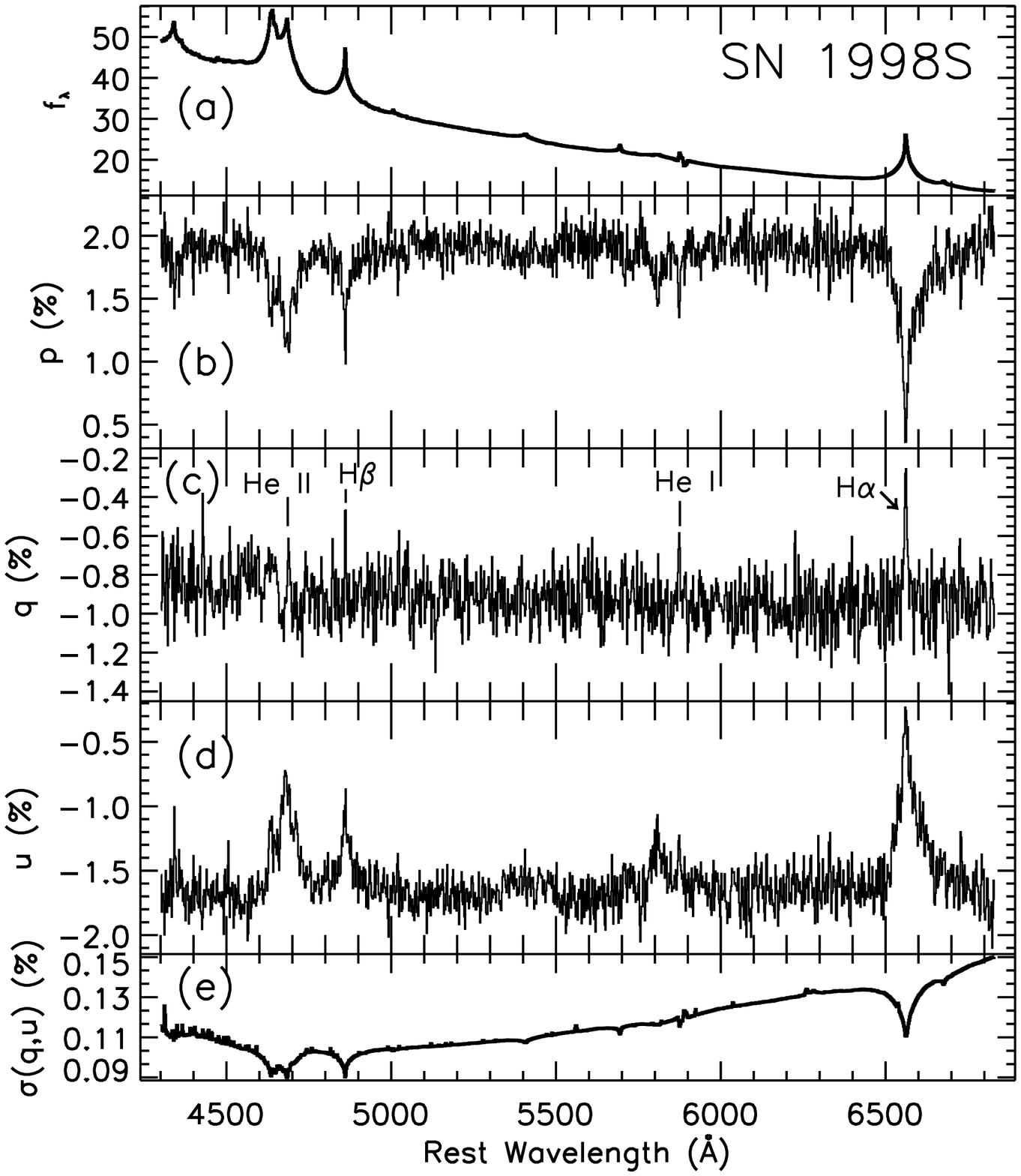,height=5truein,width=4truein,angle=0}
}
\bigskip
\noindent
{\it Figure 9:} Polarization data for SN 1998S, obtained with Keck-II on 
March 7, 1998. {\it (a)} Total flux, in units of $10^{-15}$ ergs s$^{-1}$
cm$^{-2}$ \AA$^{-1}$. {\it (b)} Observed degree of polarization. {\it (c,d)} The
normalized $q$ and $u$ Stokes parameters, with prominent narrow-line
features indicated. {\it (e)} Average of the (nearby identical) $1\sigma$
statistical uncertainties in the Stokes $q$ and $u$ parameters. See
reference \cite{leo00b} for details.\\

\bigskip
  
  We have completed our analysis of the peculiar Type IIn SN 1998S
\cite{leo00b}.  The data consist of one epoch of spectropolarimetry (5 days
after discovery) and total flux spectra spanning the first 494 days after
discovery. The SN is found to exhibit a high degree of linear polarization
(Figure 9), implying significant asphericity for its continuum-scattering
environment.  Prior to removal of the interstellar polarization, the
polarization spectrum is characterized by a flat continuum (at $p \approx 2\%$)
with distinct changes in polarization associated with both the broad
(symmetric, half width near-zero intensity $\gtrsim 10,000$ km s$^{-1}$) and
narrow (unresolved, FWHM $< 300$ km s$^{-1}$) line emission seen in the total
flux spectrum.  When analyzed in terms of a polarized continuum with
unpolarized broad-line recombination emission, however, an intrinsic continuum
polarization of $p \approx 3\%$ results, suggesting a global asphericity of
$\gtrsim 45\%$ from the oblate, electron-scattering dominated models of
H\"{o}flich \cite{hof91}.  The smooth, blue continuum evident at early times is
inconsistent with a reddened, single-temperature blackbody, instead having a
color temperature that increases with decreasing wavelength.  Broad
emission-line profiles with distinct blue and red peaks are seen in the total
flux spectra at later times, suggesting a disk-like or ring-like morphology for
the dense ($n_e \approx 10^7 {\rm\ cm^{-3}}$) circumstellar medium, generically
similar to what is seen directly in SN 1987A, although much denser and closer
to the progenitor in SN 1998S.

  The Type IIn SN 1997eg also exhibits considerable polarization \cite{leo00a};
there are sharp polarization changes across its strong, multi-component
emission lines, suggesting distinct scattering origins for the different
components. Based on our rather small sample, it appears as though SNe~II-P are
considerably less polarized than SNe~IIn, at least within the first month or
two after the explosion. Leonard et al. \cite{leo00a} show some
spectropolarimetric evidence of asphericity in the ejecta of SN II-P 1997ds,
but it does not match the degree of polarization of SNe~IIn 1998S and 1997eg.
Moreover, SN II-P 1999em does not reveal significant polarization variation
across the strong Balmer lines shortly after its explosion \cite{leo99}.

\section*{ SUPERNOVAE ASSOCIATED WITH GAMMA-RAY BURSTS?}

  At least a small fraction of gamma-ray bursts (GRBs) may be associated with
nearby SNe. Probably the most compelling example thus far is that of SN 1998bw
and GRB 980425 \cite{gal98,iwa98,woo99}, which were temporally and spatially
coincident. SN 1998bw was, in many ways, an extraordinary SN; it was very
luminous at optical and radio wavelengths, and it showed evidence for
relativistic outflow. Its bizarre optical spectrum is often classified as that
of a SN~Ic, but the object should be called a ``peculiar SN~Ic" if not a
subclass of its own; the spectrum was distinctly different from that of a
normal SN~Ic.

  As discussed by several speakers at this meeting, models suggest that SNe
associated with GRBs are highly asymmetric. Thus, spectropolarimetry should
provide some useful tests. In particular, perhaps objects such as SN 1998S,
discussed above, would have been seen as GRBs had their rotation axis been
pointed in our direction. That of SN 1998S was almost certainly {\it not}
aligned with us \cite{leo00b}; both the spectropolarimetry and the appearance
of double-peaked H$\alpha$ emission suggest an inclined view, rather than
pole-on.

\bigskip

\hbox{
\hskip +0.3truein
\psfig{figure=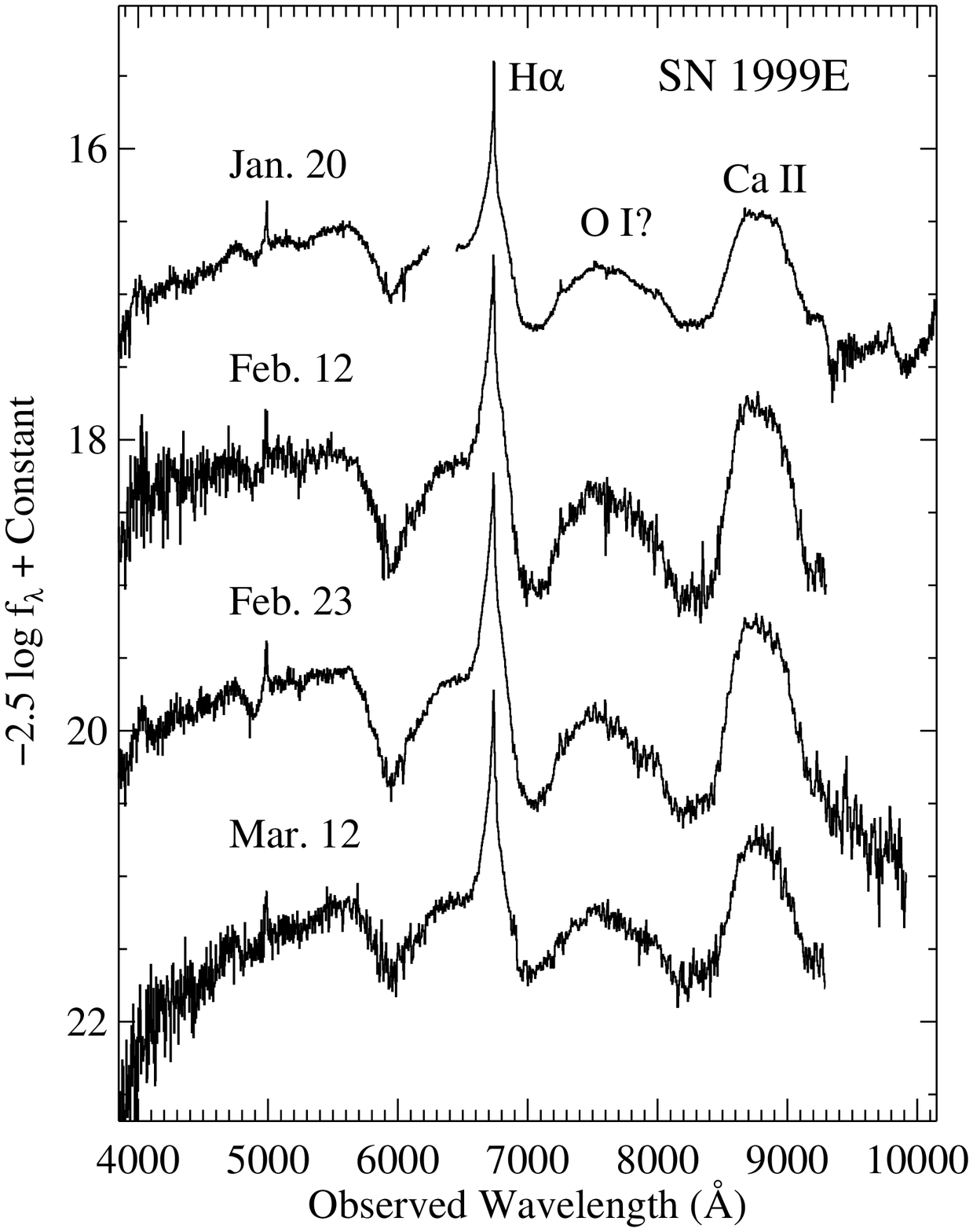,height=5truein,width=4.5truein,angle=0}
\hskip +.3truein
}
\noindent
{\it Figure 10:} Spectral evolution of SN 1999E, which may have been
associated with GRB 980910.
\medskip

   The case of GRB 970514 and the very luminous SN IIn 1997cy is also
interesting \cite{ger00,tur00}; there is a reasonable possibility that the two
objects were associated. The optical spectrum of SN 1997cy was highly unusual,
and bore some resemblance to that of SN 1998bw, though there were some
differences as well. SN 1999E, which might be linked with GRB 980910 but with
large uncertainties \cite{tho99}, also had an optical spectrum similar to that
of SN 1997cy \cite{avf99,cap99}; see Figure 10.  The undulations are very
broad, indicating high ejection velocities. Besides H$\alpha$, secure line
identifications are difficult, though some of the emission features seem to be
associated with oxygen and calcium. Perhaps SN 1999E was produced by the highly
asymmetric collapse of a carbon-oxygen core.

   Shortly before this meeting, SN~IIn 1999eb was discovered with KAIT
\cite{mod99}, and Terlevich et al. \cite{ter99} pointed out that it might be
associated with GRB 991002. However, KAIT data show that the optical SN was
visible at least 10 days {\it before} the GRB occurred, making it very unlikely
that the two were linked.  If SN 1999eb ends up showing double-peaked H$\alpha$
emission at late times, as did SN 1998S, it will be another argument against
the SN/GRB association in this particular case, since our view will not have
been pole-on.

\section*{ THE EXPANDING PHOTOSPHERE METHOD}

   Despite {\it not} being anything like ``standard candles," SNe~II-P (and
some SNe~II-L) are very good distance indicators. They are, in fact, ``custom
yardsticks" when calibrated with the ``Expanding Photosphere Method" (EPM); see
\cite{eas96}. A variant of the famous Baade-Wesselink method for determining
the distances of pulsating variable stars, this technique relies on an accurate
measurement of the photosphere's effective temperature and velocity during the
plateau phase of SNe~II-P.

  Briefly, here is how EPM works. The radius ($R$) of the photosphere can be
determined from its velocity ($v$) and time since explosion ($t - t_0$) if the
ejecta are freely expanding: $R = v(t - t_0) + R_0 \approx v(t - t_0)$,
where we have assumed that the initial radius of the star ($R_0$ at $t =
t_0$) is negligible relative to $R$ after a few days. The velocity of
the photosphere is determined from measurements of the wavelengths of the
absorption minima in P-Cygni profiles of weak lines such as those of Fe~II
or, better yet, Sc~II. (The absorption minima of strong lines like
H$\alpha$ form far above the photosphere.) The angular size ($\theta$) of
the photosphere, on the other hand, is found from the measured, dereddened
flux density ($f_\nu$) at a given frequency. We have $4 \pi D^2 f_\nu = 
4 \pi R^2 \zeta^2 \pi B_\nu (T)$, so

\[ \theta = {R\over D} = \left[{f_\nu \over {\zeta^2 \pi 
B_\nu (T)}}\right]^{1/2}, \]
 
\noindent
where $D$ is the distance to the supernova, $B_\nu (T)$ is the value 
of the Planck function at color temperature $T$ (derived from broadband
measurements of the supernova's brightness in at least two passbands),
and $\zeta^2$ is the flux dilution correction factor (basically a measure
of how much the spectrum deviates from that of a blackbody, due primarily
to the electron-scattering opacity). 

  The above two equations imply that $t = D(\theta/v) + t_0.$ Thus, for a
series of measurements of $\theta$ and $v$ at various times $t$, a plot of
$\theta/v$ versus $t$ should yield a straight line of slope $D$ and intercept
$t_0$. This determination of the distance is independent of the various
uncertain rungs in the cosmological distance ladder; it does not even depend on
the calibration of the Cepheids. It is equally valid for nearby and distant
SNe~II-P.

   An important check of EPM is that the derived distance be {\it constant}
while the SN is on the plateau (before it has started to enter the nebular
phase).  This has been verified with SN 1987A \cite{eas89} and a number of
other SNe~II-P \cite{schmidt92,schmidt94a}. Moreover, the EPM distance to SN
1987A agrees with that determined geometrically through measurements of the
brightening and fading of emission lines from the inner circumstellar ring
\cite{pan91}. It
is also noteworthy that EPM is relatively insensitive to reddening: an
underestimate of the reddening leads to an underestimate of the color
temperature $T$ [and hence of $B_\nu (T)$ as well], but this is compensated by
an underestimate of $f_\nu$, yielding a nearly unchanged value of
$\theta$. Indeed, for errors in $A_V$ [the visual extinction, or
$\sim 3.1 E(B-V)$] of 0--1 mag, one incurs an error in $D$ of only
$\sim 0$--20\% \cite{schmidt92}.

   Of course, EPM has some caveats or potential limitations. A critical
assumption is spherical symmetry for the expanding ejecta, yet polarimetry
shows that SN 1987A was not spherical \cite{jef91}, as do direct {\it HST}
measurements of the shape of the ejecta. As discussed above, the few other
SNe~II-P that have been studied don't show very much polarization, though it is
possible that deviations from spherical symmetry could be a severe problem for
some SNe~II-P. On the other hand, the {\it average} distance derived with EPM
for many SNe~II-P might be almost unaffected, given random orientations to the
line of sight. (Sometimes the cross-sectional area will be too large, and other
times too small, relative to spherical ejecta.) Indeed, comparison of EPM and
Cepheid distances to the same galaxies shows agreement to within the expected
uncertainties for a sample of 6 objects ($D_{\rm Cepheids}/D_{\rm EPM} = 0.98
\pm 0.08$; \cite{eas96}).

   Another limitation of EPM is that one needs a well-observed SN~II-P in a
given galaxy in order to measure its distance; thus, the technique is most
useful for aggregate studies of galaxies, rather than for distances of specific
galaxies in a random sample. Finally, knowledge of the flux dilution correction
factor, $\zeta^2$, is critical to the success of EPM.  Fortunately, an
extensive grid of models \cite{eas96} shows that the value of $\zeta^2$ is
mainly a function of $T$ during the plateau phase of SNe~II-P; it is relatively
insensitive to other variables such as helium abundance, metallicity, density
structure, and expansion rate. Also, it does not differ too greatly from unity
during the plateau. There are, however, some differences of opinion regarding
the treatment of radiative transfer and thermalization in expanding supernova
atmospheres \cite{bar96}.  The calculations are difficult and various
assumptions are made, possibly leading to significant systematic errors.

   The most distant SN~II-P to which EPM has successfully been applied is SN
1992am at $z = 0.0487$ \cite{schmidt94b}. The derived distance is $D =
180^{+30}_{-25}$ Mpc. This object, together with 15 other SNe~II-P at smaller
redshifts, yields a best fit value of $H_0 = 73 \pm 7$ km s$^{-1}$ Mpc$^{-1}$,
where the quoted uncertainty is purely statistical \cite{schmidt94a,eas96}. A
systematic uncertainty of $\sim \pm 6$ km s$^{-1}$ Mpc$^{-1}$ should also be
associated with the above result.  The main source of statistical uncertainty
is the relatively small number of SNe~II-P in the EPM sample, and the low
redshift of most of the objects (whose radial velocities are substantially
affected by peculiar motions).  My group is currently trying to remedy the
situation with EPM measurements of additional nearby SNe~II-P (at Lick
Observatory), as well as with Keck spectra of SNe~II-P in the redshift range
0.1--0.3.

\section*{ ACKNOWLEDGMENTS}

   My recent research on SNe has been financed by NSF grant AST--9417213, as
well as by NASA grants GO-6584, GO-7434, AR-6371, and AR-8006 from the Space
Telescope Science Institute, which is operated by AURA, Inc., under NASA
Contract NAS5-26555. The Committee on Research (U.C. Berkeley) provided partial
travel support to attend this meeting. I am grateful to the students and
postdocs who have worked with me on SNe over the past 14 years for their
assistance and discussions. Tom Matheson and Doug Leonard were especially
helpful with the figures for this review.

\end{document}